# Financial resilience of agricultural and food production companies in Spain: A compositional cluster analysis of the impact of the Ukraine-Russia war (2021-2023)

**Mike Hernandez-Romero**

*Department of Economics. Universitat de Girona, Girona, Spain*

**Germà Coenders**

*Department of Economics. Universitat de Girona, Girona, Spain*

*Corresponding author. Facultat d'Economia i Empresa. C. Universitat de Girona 10. 17003 Girona, Spain. germa.coenders@udg.edu*

*https://orcid.org/0000-0002-5204-6882*

## ABSTRACT

This study analyses the financial resilience of agricultural and food production companies in Spain amid the Ukraine-Russia war using cluster analysis based on financial ratios. This research utilizes centred log-ratios to transform financial ratios for compositional data analysis. The dataset comprises financial information from 1197 firms in Spain's agricultural and food sectors over the period 2021-2023. The analysis reveals distinct clusters of firms with varying financial performance, characterized by metrics of solvency and profitability. The results highlight an increase in resilient firms by 2023, underscoring sectoral adaptation to the conflict's economic challenges. These findings together provide insights for stakeholders and policymakers to improve sectorial stability and strategic planning.

## 1. Introduction

Tensions between Ukraine and Russia stem from a complex interplay of historical, political, and cultural factors, notably following Ukraine's independence from the Soviet Union in 1991 and Russia's annexation of Crimea in 2014 (Bielieskov and Szeligowski, 2024). The situation escalated dramatically on February 24, 2022, when Russia initiated a full-scale invasion of Ukraine. The war has had far-reaching consequences for global food supply and has prompted discussions on food security, as both Ukraine and Russia account for 30 % of the global grain export (FAO, 2023).

In addition to these direct impacts, the war has triggered cascading consequences that have amplified the ongoing global food crisis. These include labour shortages due to conscription and population displacement, disruptions in planting and harvesting cycles, and skyrocketing fertilizer prices that threaten to reduce agricultural yields (Ben Hassen and El Bilali, 2022). The rising food prices and the shift in global trade patterns have left many vulnerable economies struggling to maintain access to vital food supplies.

In this context, Spain's agricultural and food production sectors play a vital role in the country's economy, contributing significantly to the country's GDP (8.6 % in 2025) and employment (6.0 % in 2025). The sector is characterized by a diverse range of products, including fruits, vegetables, cereals, and livestock (Escudero et al., 2022). Notably, Spain is the largest worldwide producer of olive oil and the third largest of wine and ranks among the top exporters of fresh fruits and vegetables within the European Union (European Commission, DG Agriculture and Rural Development, 2024). Spain not only supplies the European Union but also relies on imports for critical agricultural inputs. The geopolitical instability has amplified the challenges in maintaining consistent food supply chains, with price volatility affecting everything from raw materials to consumer products.

In this study, specific sectors within Spain's agricultural and food production are analysed, focusing on activities classified under the Statistical Classification of Economic Activities in the European Community (NACE) 2009 codes: cereal cultivation (excluding rice), legumes and oilseeds (code 0111); the production of vegetable and animal oils and fats (code 104); milling products, starches, and starch products (code 106); and bakery and farinaceous products manufacturing (code 107).

For the cereal cultivation sector (NACE 0111), Teixeira Da Silva et al. (2023) highlight the significant disruptions caused by the closure of the Black Sea and Azov Sea ports, which severely impacted global grain exports. This disruption, coupled with rising fuel and fertilizer prices, placed a heavy strain on supply chains worldwide, including Spain's. Adding to this, Malik et al.'s (2024) findings offer a complementary perspective, particularly on the role of small businesses. In Ukraine, small enterprises in cereal, legume, and sunflower production demonstrated resilience during wartime due to their agility and compact management systems. For Spain, this could be an insightful parallel. Milling and starch products (code 106); and bakery and farinaceous products manufacturing (code 107) use cereals as raw materials.

Glauber et al. (2023) explain that vegetable oil prices—especially sunflower oil (included in code 104)—soared by over 40 % following the invasion, and the sector is now grappling with additional pressures from biofuel production and export restrictions. These factors have exacerbated supply-demand imbalances, driving prices even higher.

Spain is also significantly affected by the disruptions in Ukrainian maize production and exports. The country was expected to receive 1.9 million metric tons of maize in 2022, primarily intended as animal feed. This maize is critical for the livestock sector, which relies on high-quality feed to sustain meat and dairy production levels (Jagtap et al., 2022). With these exports disrupted by the conflict, Spain faces a rise in animal feed costs and a potential supply shortage in the domestic market. While other major producers, such as the United States, may partially compensate for this shortfall, the financial impact is inevitable due to increased transportation costs and heightened pressure on already-strained global cereal markets.

Given these challenges, understanding the financial resilience of Spain's agricultural and food production sectors is crucial. To address this, a *cluster analysis* is employed, a widely used technique for categorizing heterogeneous data into homogeneous groups, which helps identify patterns and trends within the dataset. Clustering enables the analysis of firms with similar financial characteristics, facilitating a clearer understanding of varying performance levels within a sector. The objective is to achieve high internal cohesion within each group while ensuring distinct separation between groups (Capece et al., 2010). In financial statement analysis, clustering is particularly valuable for distinguishing firms based on their financial health and resilience, empowering stakeholders to make more informed decisions about investments, risk management, and strategic planning (Caruso et al., 2018). To perform this clustering, *financial ratios* are a key instrument for assessing firm performance. These ratios provide a snapshot of various aspects of financial health, such as profitability, solvency, liquidity, and operational efficiency (Amat and Manini, 2017; Arimany-Serrat et al., 2017; Cavero Rubio et al., 2021; Krylov, 2018; Saleh et al., 2023; Tascón et al., 2018).

However, when using standard financial ratios in clustering, several challenges arise. Financial ratios often suffer from non-linearity (Carreras-Simó and Coenders, 2021; Cowen and Hoffer, 1982), asymmetry (Frecka and Hopwood, 1983; Iotti et al., 2024a; 2024b; Linares-Mustarós et al., 2018), outliers (Deshpande, 2023; Lev and Sunder, 1979) and the mutual redundancy of ratios that measure overlapping concepts (Chen and Shimerda, 1981; Linares-Mustarós et al., 2018).These issues can distort clustering results, leading to poor representation of the firms' financial profiles and the risk of forming very small clusters and even clusters composed by just one or two outliers (Dao et al., 2024; Feranecová and Krigovská, 2016; Jofre-Campuzano and Coenders, 2022; Linares-Mustarós et al., 2018; Molas-Colomer et al., 2024; Sharma et al., 2016). To overcome these limitations, this study employs the *Compositional Data* (CoDa) methodology (Aitchison, 1982; 1983; 1986; Pawlowsky-Glahn et al., 2015). This approach addresses many of the challenges posed by standard financial ratios. By means of suitable transformations, CoDa minimizes the impact of outliers, non-linearity, asymmetry and redundancy (Coenders, 2025; Coenders and Arimany-Serrat, 2025b; Dao et al., 2024; Jofre-Campuzano and Coenders, 2022; Linares-Mustarós et al., 2018; Magrini, 2025).

The application of CoDa in clustering has been widely studied since Linares-Mustarós et al. (2018) and has proven effective in various sectors. For example, Saus-Sala and colleagues used CoDa to identify clusters of firms in the farm-tourism sector based on leverage, margin, and turnover ratios (Saus-Sala et al., 2021; 2023; 2024). Similarly, Jofre-Campuzano and Coenders (2022) applied CoDa to automotive fuel companies in Spain, uncovering distinct financial profiles, including a cluster characterized by financial distress. Along the same lines, Coenders (2025) and Coenders and Arimany-Serrat (2025a) classify wine producing firms.

Arimany-Serrat and Sgorla (2024), Arimany-Serrat and Coenders (2025a), Dao et al. (2024) and Saus-Sala et al. (2024) draw clusters which are related to financial resilience during the COVID19 pandemic in the brewing, beekeeping, fishery, and farm tourism sectors, respectively. These are the articles most related to the one presented here, but none of them studies the effects of the Ukraine-Russia war.

In this study, CoDa clustering allows for the classification of firms in the agricultural and food sectors of Spain based on their financial health during the geopolitical crisis related to the Ukraine-Russia war in the hope of finding at least one cluster of financially resilient firms. These findings may serve as a useful guide for stakeholders and policymakers when considering policies aimed at enhancing the resilience of the sector in future crises.

This article is structured as follows: following the introduction, Section 2 presents a review of relevant literature and theoretical insights. Section 3 outlines the methodology, detailing the CoDa approach to clustering financial ratios and the rationale behind its use in this study. After a data description, Section 4 presents the results of the cluster analysis, highlighting the financial resilience patterns. Section 5 offers a discussion of the findings, their implications, and limitations.

## 2. Literature review

As shown by Arimany-Serrat et al. (2023) in their analysis of Spanish wineries, companies respond differently to economic shocks based on various financial and structural factors. The study revealed that during the COVID-19 lockdown, many wineries experienced a decline in margins and turnover, largely due to reduced sales. However, larger wineries with pre-existing subsidies were better able to weather the downturn, supported by stronger pre-pandemic financials. This disparity in resilience underlines the varied impact that economic shocks can have within a sector, leading to clusters of firms with differing performance outcomes. Similarly, Dao et al. (2024)'s study on Vietnamese fisheries and food production sectors identified distinct clusters, showing that those with lower leverage and higher profitability performed best. Meanwhile, Arimany-Serrat and Sgorla (2024) analysed the brewing sector in Spain and Italy during 2019–2021, revealing resilience through gradual recovery despite initial declines in profitability. Their findings emphasized sectoral heterogeneity, with distinct clusters of underperforming SMEs and high-performing corporations, and linked financial performance to transparency practices. Similar results are found by Arimany-Serrat and Coenders (2025a) in the beekeeping sector.

Based on this, within Spain's agricultural and food production sectors, the firms impacted by the Ukraine-Russia war will likely form distinct clusters. Firms with different levels of financial resilience may diverge in their ability to sustain performance and/or financial stability.

Sharif et al. (2020) highlight how major crises, such as COVID-19, drive significant volatility and uncertainty across markets. The findings show that pandemic-related shocks, heightened geopolitical risks, and economic policy uncertainty impact the U.S. stock market and alter firms' financial strategies. Baixauli-Soler et al. (2024) study the impact of such crises on the firms' optimal debt structure. Additionally, research by Chiang (2022) indicates that geopolitical risks, including the Ukraine-Russia war, significantly disrupt financial markets and corporate performance globally. Building on this, we expect that firms in the agricultural and food production sectors in Spain shift to lower-performing clusters during the 2021-2023 period due to the economic instability caused by the Ukraine-Russia war.

Much research shows that firms can recover and return to higher performance levels over time as they adapt to crisis conditions (Cirera et al., 2021; Kong et al., 2022). Cirera's findings (2021) highlight that, while firms initially face steep declines, many gradually improve by adopting digital technologies, strengthening managerial practices, and diversifying into new markets. Similarly, Boungou and Yatié (2022) underscore the immediate negative impact of the Ukraine-Russia war on global stock markets, especially in countries geographically closer to the conflict or strongly condemning the invasion, but with some evidence of partial recovery over time.

These findings suggest that adaptation processes, supported by strategic realignments and resilience measures, enable firms to mitigate adverse impacts and eventually reposition themselves within higher-performing clusters. Based on these findings, we expect some firms to shift back to better-performing clusters as they adapt to the crisis, as shown in similar research by Arimany-Serrat and Sgorla (2024), Arimany-Serrat and Coenders (2025a), Dao et al. (2024), and Saus-Sala et al. (2024).

## 3. Methods

This section outlines the methodology employed in this study, focusing on the CoDa approach and how it is applied for clustering financial data of agricultural and food production companies in Spain.

The financial data used in this study was obtained from the Sistema de Análisis de Balances Ibéricos (SABI, accessible at https://sabi.bvdinfo.com) database. The focus was placed on companies classified under the following primary codes of the NACE 2009: the cultivation of cereals (excluding rice), legumes, and oilseeds (code 0111); the manufacture of vegetable and animal oils and fats (code 104); the production of milling products, starches, and starch products (code 106); and the manufacture of bakery and farinaceous products (code 107).

In addition to sectoral classification, companies were required to have a minimum of ten employees to ensure the inclusion of firms with sufficient operational activity. Only companies registered within Spain were considered, covering various legal forms such as private and public limited companies, and other business structures.

After the initial data extraction, companies with zero values in key financial indicators such as total assets, operating income, and operating expenses for any of the three study years (2021, 2022 and 2023) were removed, as these zero values indicated periods of inactivity during the specified years. The remaining zero values were handled according to the procedures outlined in Section 3.4 and in Coenders and Arimany-Serrat (2025b).

The final dataset comprised 140 companies involved in cereal cultivation (sector 0111), 95 companies operating in the manufacture of vegetable and animal oils and fats (sector 104), 65 companies in the milling products and starches sector (sector 106), and 897 companies engaged bakery and farinaceous products (sector 107), for a total of 1197 companies and 3591 cases for the three-year period under study. These 1197 companies account for 18.1 % of the selected sectors in numbers, and 59.2 % in revenue (data for 2023).

## 3.1. Compositional financial statement analysis

In the CoDa methodology, a composition is defined as a set of $D$ strictly positive numbers (parts) where only the relative magnitude of parts to one another is of interest (Aitchison, 1982; Coenders et al., 2023; Filzmoser et al., 2018; Greenacre, 2018; Pawlowsky-Glahn et al., 2015; Van den Boogaart and Tolosana-Delgado, 2013). In scientific fields such as chemistry, compositions often sum to a constant value, but in financial analysis this is not required:

$$\mathbf{x} = (x_1, x_2, \ldots, x_D) \text{ with } x_j > 0,\ j = 1,2,\ldots,D \quad (1)$$

There are two main rules for applying the CoDa methodology to financial data (Coenders and Arimany-Serrat, 2025b; Creixans-Tenas et al., 2019):

- Avoiding negative values, as they can lead to misinterpretation and discontinuities in ratios (e.g., a positive return on equity when both profit and net worth are negative).
- Avoiding overlapping parts. For example, using both total assets and non-current assets is problematic, as the latter is part of the former. Only the full amalgamation (e.g., total assets) or the individual parts (e.g., non-current and current assets) should be used, not both.

In this article, the $D = 6$ positive and non-overlapping financial statement categories $x_j$ were selected to align with this study's focus on liquidity, solvency, operational efficiency, and profitability (Creixans-Tenas et al., 2019; Coenders and Arimany-Serrat, 2025a). Excessive categorization into a high number $D$ of accounting figures risks data sparsity, such as frequent zero value entries, particularly in samples containing small firms. The first four categories correspond to balance sheet items, while the last two are aggregate values from the profit and loss account:

- $x_1$=non-current assets,
- $x_2$=current assets,
- $x_3$=non-current liabilities,
- $x_4$=current liabilities,
- $x_5$=revenue (net sales),
- $x_6$=expenses (operating expenses).

The ratios used in this study are those outlined by in Arimany-Serrat and Coenders (2025a), Dao et al. (2024), Jofre-Campuzano and Coenders (2022), and Saus-Sala et al. (2024) in articles clustering firms according to their financial resilience in front of crises. They are based on the previous six key financial-statement figures and include:

Turnover ratio:

$$x_5 / (x_1 + x_2). \quad (2)$$

Current-asset turnover ratio:

$$x_5 / x_2. \quad (3)$$

Profit margin ratio:

$$(x_5 - x_6) / x_5. \quad (4)$$

Leverage ratio:

$$(x_1 + x_2) / (x_1 + x_2 - x_3 - x_4). \tag{5}$$

Return on Assets (ROA):

$$(x_5 - x_6) / (x_1 + x_2), \tag{6}$$

which can also be derived by multiplying the margin by the turnover ratio.

Return on Equity (ROE):

$$(x_5 - x_6) / (x_1 + x_2 - x_3 - x_4), \tag{7}$$

which can also be obtained by multiplying the ROA by the leverage ratio.

Debt ratio:

$$(x_3 + x_4) / (x_1 + x_2). \tag{8}$$

Short-term debt ratio:

$$x_4 / (x_1 + x_2). \tag{9}$$

Long-term solvency ratio:

$$(x_1 + x_2) / (x_3 + x_4). \tag{10}$$

Short-term solvency ratio a.k.a. liquidity ratio and current ratio:

$$x_2 / x_4. \tag{11}$$

Asset tangibility ratio:

$$x_1 / x_2. \tag{12}$$

Debt maturity ratio:

$$x_3 / x_4. \tag{13}$$

## 3.2. Centered log-ratios

To apply the CoDa methodology, financial data must first be transformed into a Euclidean space using log-ratio transformations, a standard approach in CoDa analysis (Aitchison, 1986). The specific transformation known as centred log-ratios (CLR) (Aitchison, 1983) retains the relative distances between data points, thereby enabling the use of traditional distance-based statistical methods, such as cluster analysis, on the transformed data. The CLR transformation for each accounting figure in Equation 1 is defined as:

$$CLR_j = \log\left(\frac{x_j}{\sqrt[D]{x_1 x_2 \dots x_D}}\right) \quad \text{with } j = 1,2,\dots,D. \tag{14}$$

In financial statement analysis, this transformation involves comparing each accounting figure $x_j$ to the geometric mean of all for a given firm. This transformation solves the challenges of non-linearity, asymmetry, and outliers encountered in traditional financial ratios (Arimany-Serrat and Coenders, 2025a; Arimany-Serrat and Sgorla, 2024; Carreras-Simó and Coenders, 2020; Saus-Sala et al., 2021; 2023; 2024).

## 3.3. Cluster analysis

Cluster analysis is a multivariate statistical technique aimed at grouping firms based on the similarity of their financial structures. In the CoDa context, this method identifies clusters of companies with comparable financial profiles, which is particularly useful for understanding sectoral resilience (Arimany-Serrat and Coenders, 2025a; Arimany-Serrat and Sgorla, 2024; Dao et al., 2024; Saus-Sala et al., 2024).

By applying CLR transformations, traditional Euclidean distances correspond to Aitchison's distances, which are the standard in CoDa (Aitchison, 1983; Aitchison et al., 2000). Consequently, the distance between two firms, $m$ and $l$, is calculated as:

$$d_{ml} = \sqrt{(CLR_{1m} - CLR_{1l})^2 + (CLR_{2m} - CLR_{2l})^2 + \cdots + (CLR_{Dm} - CLR_{Dl})^2} \quad (15)$$

This transformation solves the commonest problems of standard financial ratios in cluster analysis, which tend to form very large and very small clusters and even clusters composed by just one or two outliers (Dao et al., 2024; Feranecová and Krigovská, 2016; Jofre-Campuzano and Coenders, 2022; Linares-Mustarós et al., 2018; Molas-Colomer et al., 2024; Sharma et al., 2016). Moreover, the clusters obtained do not depend on the particular ratios chosen in Equations 2 to 13, but only on the $D$ $x_j$ financial statement categories. They are thus immune to ratio redundancy, which would give more weight to the redundant ratios in the final classification.

This approach allows for the use of common clustering algorithms in financial performance analysis, such as Ward's method (Ward jr, 1963) and k-means (MacQueen, 1967). In this study, the k-means algorithm, as implemented in CoDaPack, is applied, following the methodology described by Coenders and Arimany-Serrat (2025a; 2025b) to classify firms based on their financial profiles. Clustering is made on the pooled data over all three years jointly, as is standard practice in related research (Arimany-Serrat and Coenders, 2025a; Arimany-Serrat and Sgorla, 2024; Dao et al., 2024; Saus-Sala et al., 2023; 2024) to ensure comparability of the cluster structures over time, and the meaningfulness of firm migration from one cluster to another.

Figure 1 illustrates the evolution of both the average silhouette width (Kaufman and Rousseeuw, 1990) and the Caliński-Harabasz (Caliński and Harabasz, 1974) indices across different values of $k$ (numbers of clusters). In this case, a three-cluster solution maximizes both criteria, indicating the most appropriate grouping for the present analysis.

**Figure 1.** Evaluation of clustering solutions using the average silhouette width and the Caliński-Harabasz indices

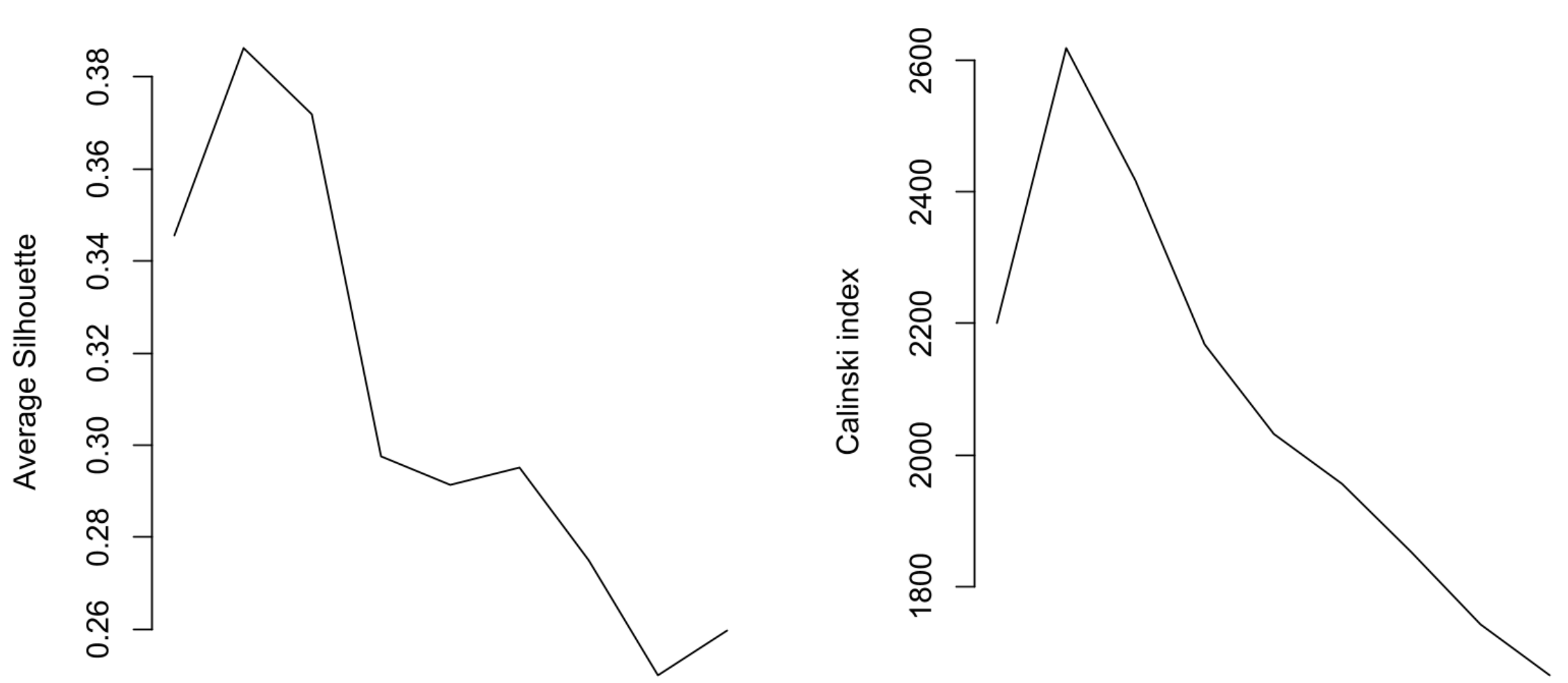

## 3.4. Handling zero values

A common limitation of both compositional and standard financial ratios is their inability to handle zero values in accounting data, as ratios involving zero are undefined (Martín-Fernández et al., 2011). However, the CoDa methodology offers advanced tools for zero imputation before log-ratio computation, which ensures valid analysis.

The most widely used imputation method in compositional financial analysis is the Expectation-Maximization (EM) algorithm for log-ratios, developed by Palarea-Albaladejo and Martín-Fernández (2008). This method predicts zero values based on the existing data under standard statistical assumptions while constraining them to be below a small value selected by the user (e.g., the 5$^{th}$ percentile of non-zero values).

For this study, only one accounting figure had zeros (non-current liabilities for 14.9 % of cases), which were imputed with the method just described. The issue is discussed in depth in Coenders and Arimany-Serrat (2025b).

The zero replacement, classification and further analyses were carried out with CoDaPack2.03.06 (Comas-Cufí and Thió-Henestrosa, 2011; Thió-Henestrosa and Martín-Fernández, 2005). CoDaPack is freely available at https://ima.udg.edu/codapack/. See Coenders and Arimany-Serrat (2023; 2025a) for an introduction to the use of CoDaPack in financial statement analysis in English and Spanish, respectively.

# 4. Results

## 4.1. Sector ratio averages

A straightforward way to use financial ratios statistically is to determine average ratios within a sector. In compositional data analysis, this is achieved by calculating the *compositional centre* (Aitchison, 1997). This centre is defined as the array of geometric means of all firms for each individual part, normalized so that the values sum to one (Table 1).

This is not to be mistaken with the geometric mean of all parts for each individual firm, which are used to compute the CLR. By employing the compositional centre, the geometric-mean approach under the CoDa framework makes it possible to compute standard sector-level financial ratios (Arimany-Serrat and Coenders, 2025a; Arimany-Serrat and Sgorla, 2024; Saus-Sala et al., 2021; 2023; 2024).

**Table 1.** Compositional centre (geometric means normalized to unit sum) for the selected agricultural and food production sectors

| | | 2021 | 2022 | 2023 |
|---|---|---|---|---|
| $x_1$ | Non-current assets | 0.0730 | 0.0712 | 0.0696 |
| $x_2$ | Current assets | 0.1237 | 0.1145 | 0.1175 |
| $x_3$ | Non-current liabilities | 0.0200 | 0.0155 | 0.0129 |
| $x_4$ | Current liabilities | 0.0728 | 0.0711 | 0.0696 |
| $x_5$ | Revenue | 0.3638 | 0.3712 | 0.3773 |
| $x_6$ | Expenses | 0.3467 | 0.3565 | 0.3531 |

One notable advantage of using geometric means is their property that the ratio of geometric means between two parts equals the geometric mean of their ratios. Let $g(x_i)$ be the geometric mean of the $i$th accounting figure over a sample of firms:

$$g(x_i / x_j) = g(x_i) / g(x_j) \tag{16}$$

This contrasts with arithmetic means, which do not share this property. Using arithmetic means of the accounting figures and then computing ratios at the sector level can produce results that contradict those obtained by first calculating ratios at the firm level and then averaging them (Saus-Sala et al., 2021).

Using the geometric mean property, the mean sectorial current-asset turnover ratio for 2021 (Equation 3) can be expressed as:

$$g(x_5) / g(x_2) = 0.3638 / 0.1237 = 2.940 \tag{17}$$

Similarly, the margin ratio (Equation 4) for the same year is:

$$(g(x_5) - g(x_6)) / g(x_5) = (0.3638 - 0.3467) / 0.3638 = 0.047 \tag{18}$$

The mean sectoral financial ratios thus computed (Table 2) reveal that firms experienced financial strain during the initial period of the war. In 2022, the ROA, and ROE ratios dropped significantly compared to 2021. Importantly, the decline in these profitability measures was not due to inefficient asset utilization or to leverage—as both the turnover ratio and the current-asset turnover ratio increased while long-term solvency remained virtually identical—but rather can be attributed to decreased margin. However, by 2023 a notable recovery was observed; these profitability measures not only returned to pre-war levels but even surpassed the 2021 figures. These findings suggest that while the initial economic shock triggered by the onset of the war set off significant structural changes, firms demonstrated remarkable resilience, ultimately emerging in a stronger and more financially efficient position.

**Table 2.** Annual mean financial ratios for the selected agricultural and food production sectors

| | 2021 | 2022 | 2023 |
|---|---|---|---|
| Turnover ratio | 1.849 | 1.998 | 2.016 |
| Current-asset turnover ratio | 2.940 | 3.241 | 3.211 |
| Profit margin ratio | 0.047 | 0.039 | 0.064 |
| Leverage ratio | 1.893 | 1.873 | 1.788 |
| ROA | 0.086 | 0.079 | 0.129 |
| ROE | 0.164 | 0.148 | 0.231 |
| Debt ratio | 0.471 | 0.466 | 0.440 |
| Short-term debt ratio | 0.370 | 0.382 | 0.371 |
| Long-term solvency ratio | 2.120 | 2.144 | 2.268 |
| Short-term solvency ratio | 1.699 | 1.610 | 1.688 |
| Asset tangibility ratio | 0.590 | 0.621 | 0.592 |
| Debt maturity ratio | 0.274 | 0.218 | 0.185 |

A second axis of analysis relates to debt structure. The debt maturity ratio experienced a consistent reduction over the three-year period. This decline was particularly pronounced from 2021 to 2022. This behaviour is also observed in the short-term debt ratio. The initial increase in 2022 suggests a reactive financial adjustment to the pressures of the war, where access to long-term debt may have been restricted. This change in the debt structure had a direct impact on liquidity. The short-term solvency ratio (liquidity ratio) declined from 2021 to 2022, suggesting a temporary deterioration in financial stability. However, from 2022 to 2023, short-term solvency showed a recovery. This is associated with the rebound in key profitability indicators (margin, ROA, and ROE), as well as improved turnover ratios. In other words, while the war initially weakened firms' solvency due to their increased reliance on short-term debt and declining profitability, the financial recovery in 2023 appears to have mitigated these risks.

Table 3 provides a more nuanced analysis by disaggregating the financial ratios on an annual basis by sector, a perspective that contrasts with the global approach of Table 2. In Table 3, the evolution of profitability ratios (profit margin, ROA, and ROE) between 2021 and 2022 reveals that sectors 106 and 0111 benefited from slight improvements in margin, ROA and ROE. This stability in profitability indicators suggests that, despite the economic shock induced by the Ukraine-Russia war, these sectors managed to maintain or even enhance their operational performance, possibly by keeping cost pressures at bay or by adapting their pricing strategies in a turbulent market environment.

In contrast, sector 107, which comprises a much larger number of companies (897 firms operating in the bakery and farinaceous products sector), was the only one to suffer a pronounced deterioration in profitability during the same period. The data indicate that from 2021 to 2022, the profit margin, ROA, and ROE ratios in sector 107 fell significantly. This decline is likely due to the competitive nature of the sector, where profit margins tend to be narrow, and even minor increases in input costs—stemming from supply chain disruptions or heightened commodity prices—can erode profitability. However, it is notable that from 2022 to 2023, sector 107 experienced the largest rebound in these ratios, a classic rebound effect: the deeper the initial fall, the more pronounced the recovery once market conditions begin to stabilize. Sector 104, representing companies involved in the manufacture of vegetable and animal oils and fats, also displayed an interesting pattern, with hardly any change in profitability between 2021 and 2022.

**Table 3.** Annual mean financial ratios by NACE code (104: oils and fats; 106: milling products and starches; 107: bakery and farinaceous products; 0111 cultivation of cereals, legumes and oilseeds)

| | Code 104 | | | Code 106 | | |
|---|---|---|---|---|---|---|
| | 2021 | 2022 | 2023 | 2021 | 2022 | 2023 |
| Turnover ratio | 1.908 | 2.160 | 1.888 | 1.902 | 2.199 | 2.278 |
| Current-asset turnover ratio | 2.003 | 2.264 | 1.978 | 2.004 | 2.310 | 2.413 |
| Profit margin ratio | 0.067 | 0.063 | 0.080 | 0.038 | 0.054 | 0.054 |
| Leverage ratio | 2.574 | 2.499 | 2.383 | 2.162 | 2.085 | 1.778 |
| ROA | 0.128 | 0.136 | 0.151 | 0.073 | 0.118 | 0.123 |
| ROE | 0.330 | 0.341 | 0.361 | 0.157 | 0.245 | 0.218 |
| Debt ratio | 0.612 | 0.600 | 0.580 | 0.537 | 0.520 | 0.437 |
| Short-term debt ratio | 0.520 | 0.526 | 0.520 | 0.459 | 0.460 | 0.387 |
| Long-term solvency ratio | 1.635 | 1.667 | 1.723 | 1.861 | 1.922 | 2.286 |
| Short-term solvency ratio | 1.832 | 1.813 | 1.836 | 2.068 | 2.069 | 2.440 |
| Asset tangibility ratio | 0.050 | 0.048 | 0.047 | 0.054 | 0.051 | 0.059 |
| Debt maturity ratio | 0.176 | 0.140 | 0.116 | 0.171 | 0.131 | 0.131 |
| | Code 107 | | | Code 0111 | | |
| | 2021 | 2022 | 2023 | 2021 | 2022 | 2023 |
| Turnover ratio | 1.816 | 1.934 | 2.014 | 0.934 | 0.962 | 0.923 |
| Current-asset turnover ratio | 3.577 | 3.954 | 3.989 | 1.305 | 1.356 | 1.273 |
| Profit margin ratio | 0.040 | 0.028 | 0.053 | 0.078 | 0.089 | 0.129 |
| Leverage ratio | 1.733 | 1.698 | 1.647 | 1.555 | 1.583 | 1.562 |
| ROA | 0.073 | 0.054 | 0.106 | 0.072 | 0.085 | 0.119 |
| ROE | 0.127 | 0.091 | 0.175 | 0.113 | 0.135 | 0.186 |
| Debt ratio | 0.423 | 0.411 | 0.393 | 0.357 | 0.368 | 0.360 |
| Short-term debt ratio | 0.326 | 0.333 | 0.328 | 0.278 | 0.299 | 0.298 |
| Long-term solvency ratio | 2.364 | 2.432 | 2.547 | 2.801 | 2.715 | 2.779 |
| Short-term solvency ratio | 1.556 | 1.469 | 1.540 | 2.571 | 2.373 | 2.429 |
| Asset tangibility ratio | 0.970 | 1.045 | 0.981 | 0.398 | 0.410 | 0.379 |
| Debt maturity ratio | 0.296 | 0.235 | 0.198 | 0.283 | 0.233 | 0.206 |

Regarding the debt structure, Table 3 corroborates the findings from Table 2 by showing that the onset of the Ukraine-Russia war prompted all sectors to shift their financing strategies toward short-term debt according to the debt-maturity ratio. However, short term solvency (liquidity) only worsened in 2022 for two sectors: 107 and 0111.

Among the studied sectors, sector 106—comprising 65 companies in the milling products and starches sector—stood out for the most significant adjustment between 2022 and 2023. This sector managed to reduce both its overall debt ratio and short-term debt ratio more than any of the others, thereby achieving the highest short-term solvency by 2023. The relatively smaller number of firms in sector 106 may have allowed for quicker, more coordinated negotiations with lenders, facilitating a rapid restructuring of their debt profiles in response to the crisis.

In summary, while the aggregated data in Table 2 might suggest a global decline in profitability following the outbreak of the war, the sectoral breakdown in Table 3 reveals that this adverse impact was largely confined to sector 107. The other sectors—104, 106, and 0111—either maintained stable profitability or even improved their financial performance over the period analysed. Additionally, although all sectors experienced a shift toward short-term debt financing, the debt restructuring undertaken by sector 106 in 2023 highlights its superior agility and financial management in a volatile economic landscape. These findings not only illuminate the differential impacts of the war across sectors but also underscore the importance of sector-specific strategies in navigating financial crises.

## 4.2. Cluster characterization

The cluster analysis identified three distinct financial profiles, each demonstrating a unique strategy for managing solvency, profitability, and operational efficiency amid economic instability. These divergent approaches, detailed in Table 4, confirm that firms adopt distinct financial strategies and have varying resilience.

**Table 4.** Financial ratios by cluster

| | Cluster 1 | Cluster 2 | Cluster 3 |
|---|---|---|---|
| Turnover ratio | 0.549 | 2.606 | 1.402 |
| Current-asset turnover ratio | 3.990 | 2.790 | 2.716 |
| Profit margin ratio | 0.042 | 0.057 | 0.048 |
| Leverage ratio | 1.200 | 9.801 | 1.276 |
| ROA | 0.023 | 0.150 | 0.067 |
| ROE | 0.027 | 1.473 | 0.086 |
| Debt ratio | 0.167 | 0.897 | 0.216 |
| Short-term debt ratio | 0.096 | 0.611 | 0.213 |
| Long-term solvency ratio | 5.985 | 1.113 | 4.619 |
| Short-term solvency ratio | 1.420 | 1.528 | 2.412 |
| Asset tangibility ratio | 6.259 | 0.070 | 0.937 |
| Debt maturity ratio | 0.722 | 0.469 | 0.012 |

Cluster 1 (34 % of observations) represents firms characterized by a heavy reliance on tangible assets. By maintaining a leverage ratio that reveals a preference for equity over debt, and a debt maturity that reveals a preference for long-term debt, these firms exhibit a conservative financial structure. Their high asset tangibility ratio suggests significant investments in non-current assets, such as machinery or property, which may indicate capital-intensive operations. However, this conservative approach comes at a cost: these firms exhibit the lowest profitability metrics, with a ROA of 0.023 and ROE of 0.028, signalling inefficiency in converting assets or equity into profits.

Cluster 2 (42 % of observations) comprises high-risk firms with extreme leverage, relying heavily on debt to finance operations. This aggressive strategy artificially inflates their ROE through leverage magnification, masking underlying vulnerabilities. However, the turnover figure leads to a satisfactory ROA. Their debt ratio reveals that nearly 90 % of assets are debt-financed. Paired with a high short-term debt ratio, it exposes them to refinancing risks and insolvency. Their asset-light structure (tangibility ratio) suggests operations dependent on volatile inputs.

The firms in Cluster 3 (24 % of observations) adopted a balanced financial strategy, combining moderate leverage with robust liquidity (short-term solvency ratio), enabling agility amid volatility. Their turnover ratio and profitability (ROA: 0.067, ROE: 0.086) are acceptable. While leveraging short-term debt (maturity ratio) for flexibility, they avoided Cluster 2's excessive risk and Cluster 1's rigidity and low profitability.

Cluster 3 is by far the best, most balanced and most resilient. It has both profitability and solvency to spare, while the other clusters are deficient in one of these two characteristics. A slight shock in prices or costs could bring the ROA and ROE of Cluster 1 into loss figures. The high leverage of Cluster 2 could translate a slightly negative margin into a large negative ROE, and, if prolonged for a few years, into bankruptcy.

## 4.3. Relation between cluster and other variables

This section explores the relationship between the three identified clusters and other variables including NACE code, legal structure, year, trade activities (imports and exports), number of employees, and region.

The mosaic plot in Figure 2 illustrates the association between the clusters and the NACE codes. The height of each bar indicates the percentage of companies from each NACE sector within a cluster, while the width of the bars represents the size of the clusters. The majority of firms are concentrated in Cluster 1 and Cluster 2, which together account for over 75 % of the sample. Cluster 3, associated to higher resilience, is the smallest.

**Figure 2.** Mosaic plot between cluster and NACE code (0111 cultivation of cereals, legumes and oilseeds; 104: oils and fats; 106: milling products and starches; 107: bakery and farinaceous products)

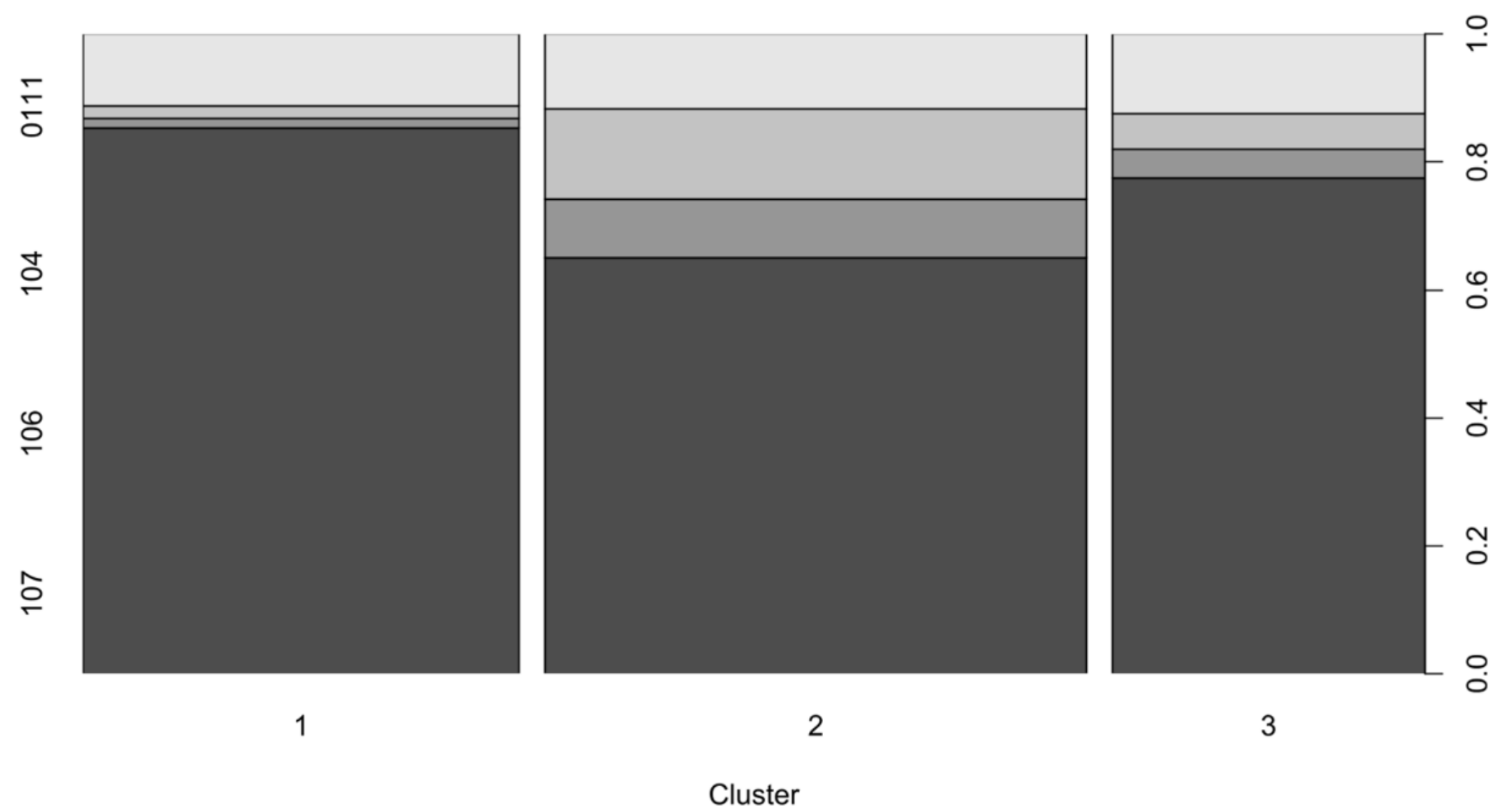

Cluster 1 is associated with sector 107 (bakery and farinaceous products), aligning with its conservative financial structure and reliance on tangible assets. There is virtually no presence of sectors 106 and 104 in Cluster 1. Cluster 2 is linked to sectors 104 and 106 (vegetable and animal oils/fats, milling and starches), reflecting its high-risk leverage strategy and profitability. Cluster 3 exhibits no clear sectoral dominance, reflecting its heterogeneous composition.

Figure 3 illustrates the relationship between the clusters and the three years of the analysed period. The number of companies in Cluster 3 increases in 2023, which indicates that as firms adapted to the prolonged effects of the conflict, they tended to shift to a better-performing cluster. However, unlike the findings in Arimany-Serrat and Sgorla (2024), Dao et al. (2024), and Saus-Sala et al. (2024), there is no shrinkage of this cluster during 2022, marking the start of the crisis.

**Figure 3.** Mosaic plot between cluster and year

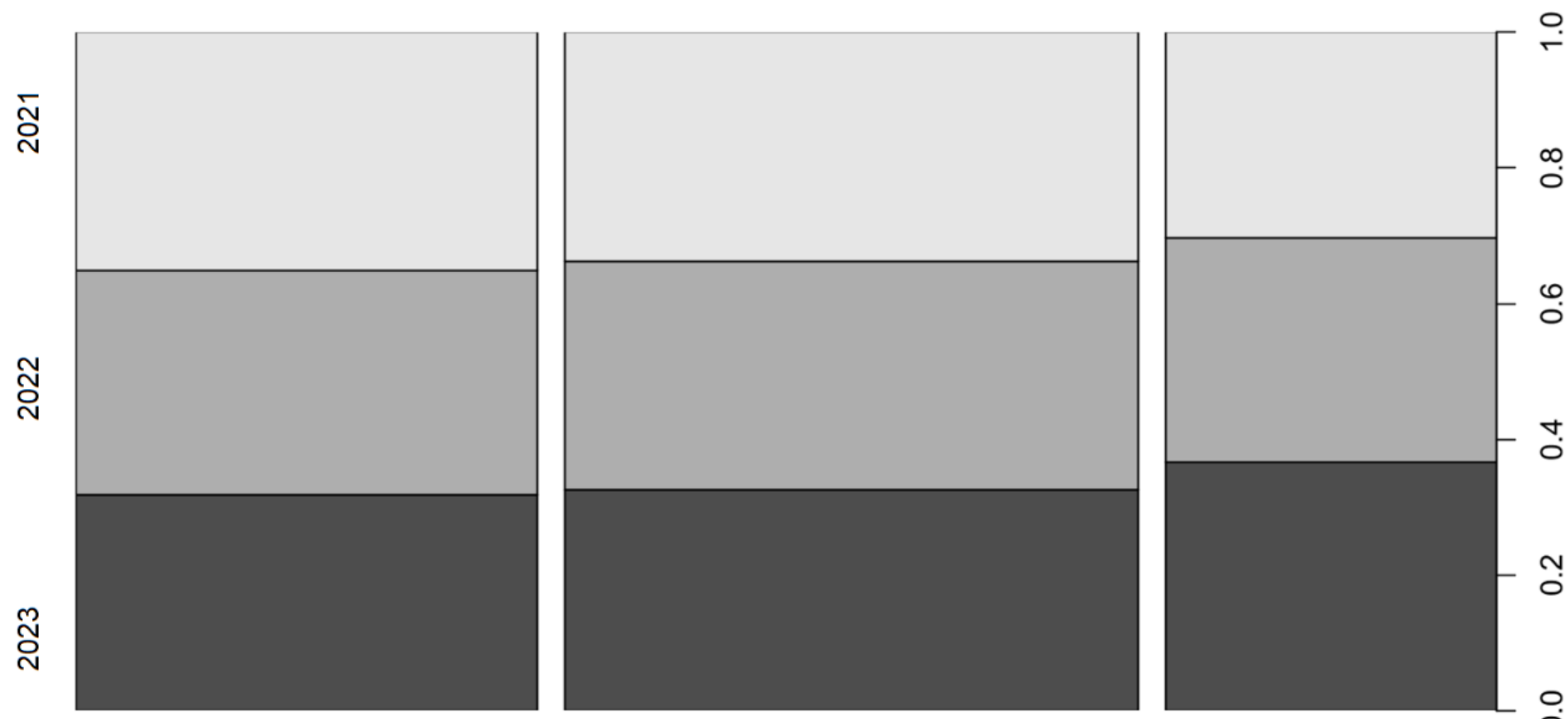

If we look at the transitions of individual firms from and to Cluster 3, altogether, the net gain for Cluster 3 is 25 firms in 2022 and 56 in 2023:

- In 2022 35 companies moved from Cluster 1 to Cluster 3 and 21 the other way around.
- In 2022 21 companies moved from Cluster 2 to Cluster 3 and 10 the other way around.
- In 2023 37 companies moved from Cluster 1 to Cluster 3 and 16 the other way around.
- In 2023 28 companies moved from Cluster 2 to Cluster 3 and 18 the other way around.

Figure 4 shows that Cluster 1 is the cluster with the lowest proportion of public limited companies and Cluster 2 is the one with the highest proportion. This may be because public companies, with greater access to capital markets and higher investor expectations, tend to adopt more aggressive, debt-driven financial strategies, resulting in the high-risk profile observed in Cluster 2. Conversely, private limited companies often follow more conservative approaches and have less access to debt, aligning them with the characteristics of Cluster 1. There are virtually no companies of other legal forms in any of the clusters.

**Figure 4.** Mosaic plot between cluster and legal form

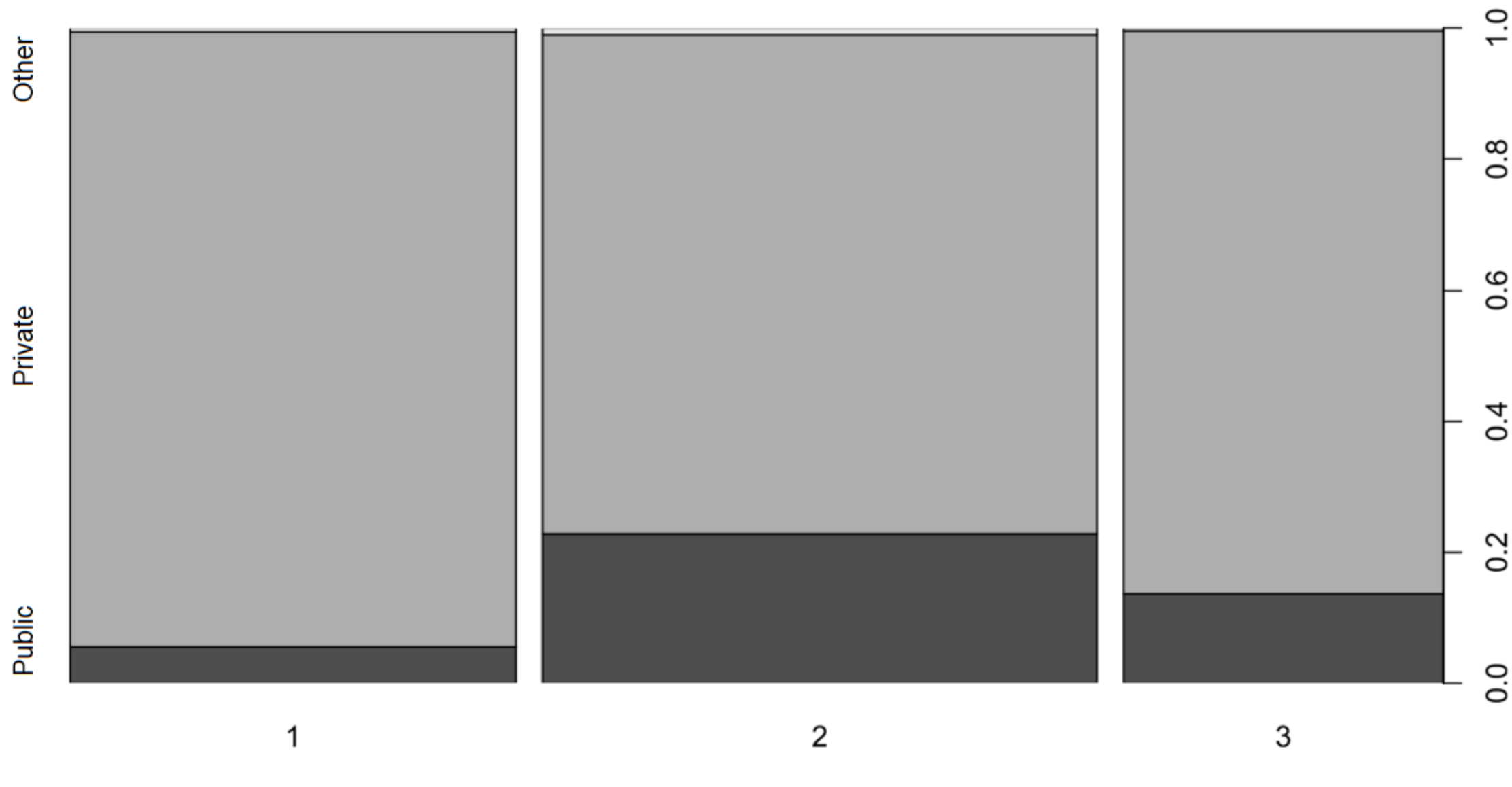

A similar conclusion can be drawn from Figure 5, which illustrates the distribution of employee numbers across clusters. Interpreting employee count as an indicator of operational capacity, it can be observed that Cluster 2 is characterized by firms with the highest number of employees. This reinforces the pattern observed in Figure 4: firms with greater operational scale, such as public limited companies with access to broader capital markets, are more likely to adopt aggressive, risk-laden financial strategies. The most resilient Cluster 3 is characterized by a small median number of employees.

**Figure 5.** Boxplots of employees by cluster

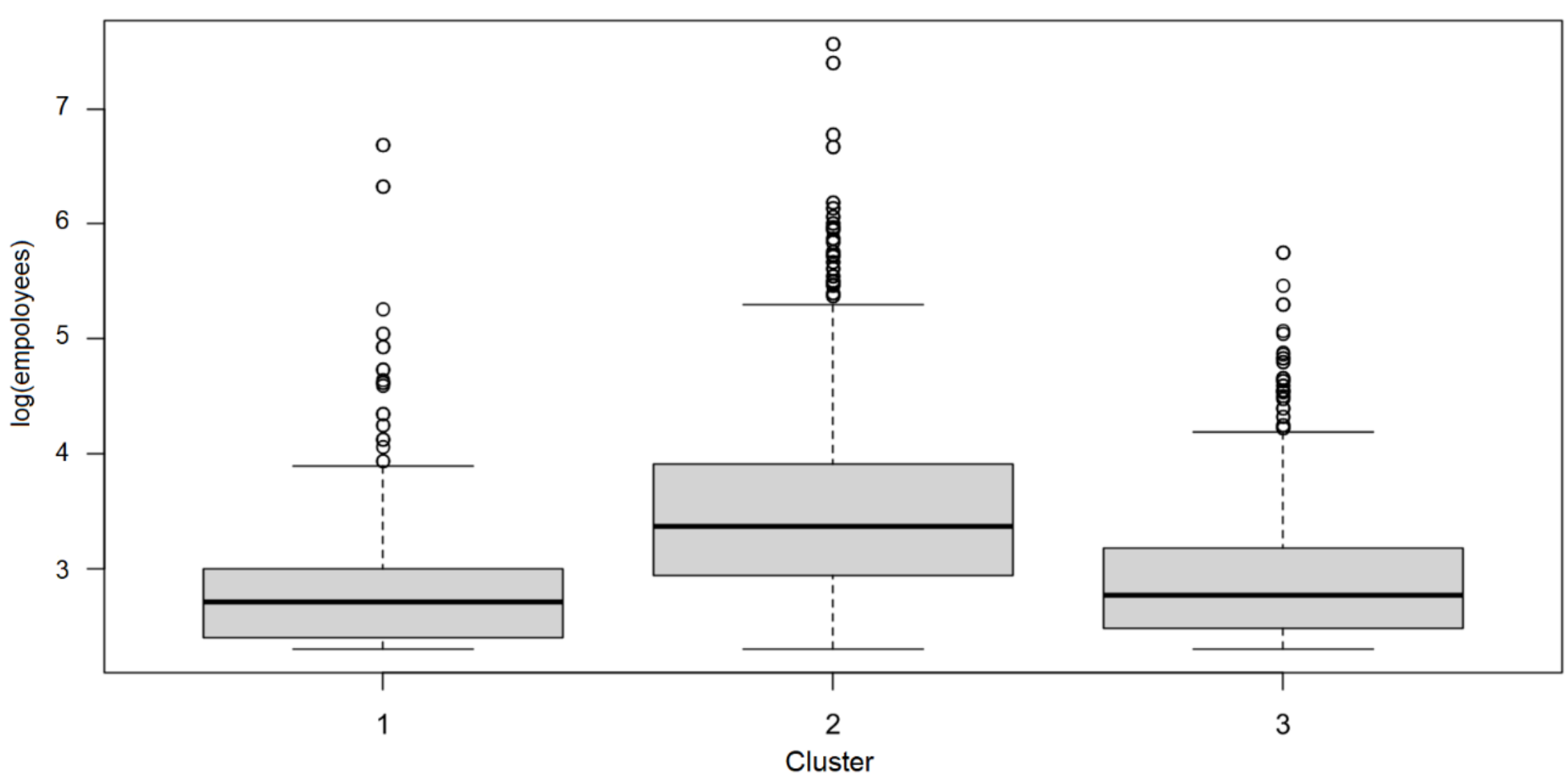

Figures 6 and 7 illustrate the association between clusters and trade activity. Cluster 1, characterized by conservative, asset-intensive firms, exhibits the highest import/export activity. In contrast, Cluster 2, comprised of high-risk, leverage-driven firms, shows the lowest trade activity; their asset-light structure and dependency on volatile inputs make them more vulnerable to trade fluctuations.

**Figure 6.** Mosaic plot between cluster and imports

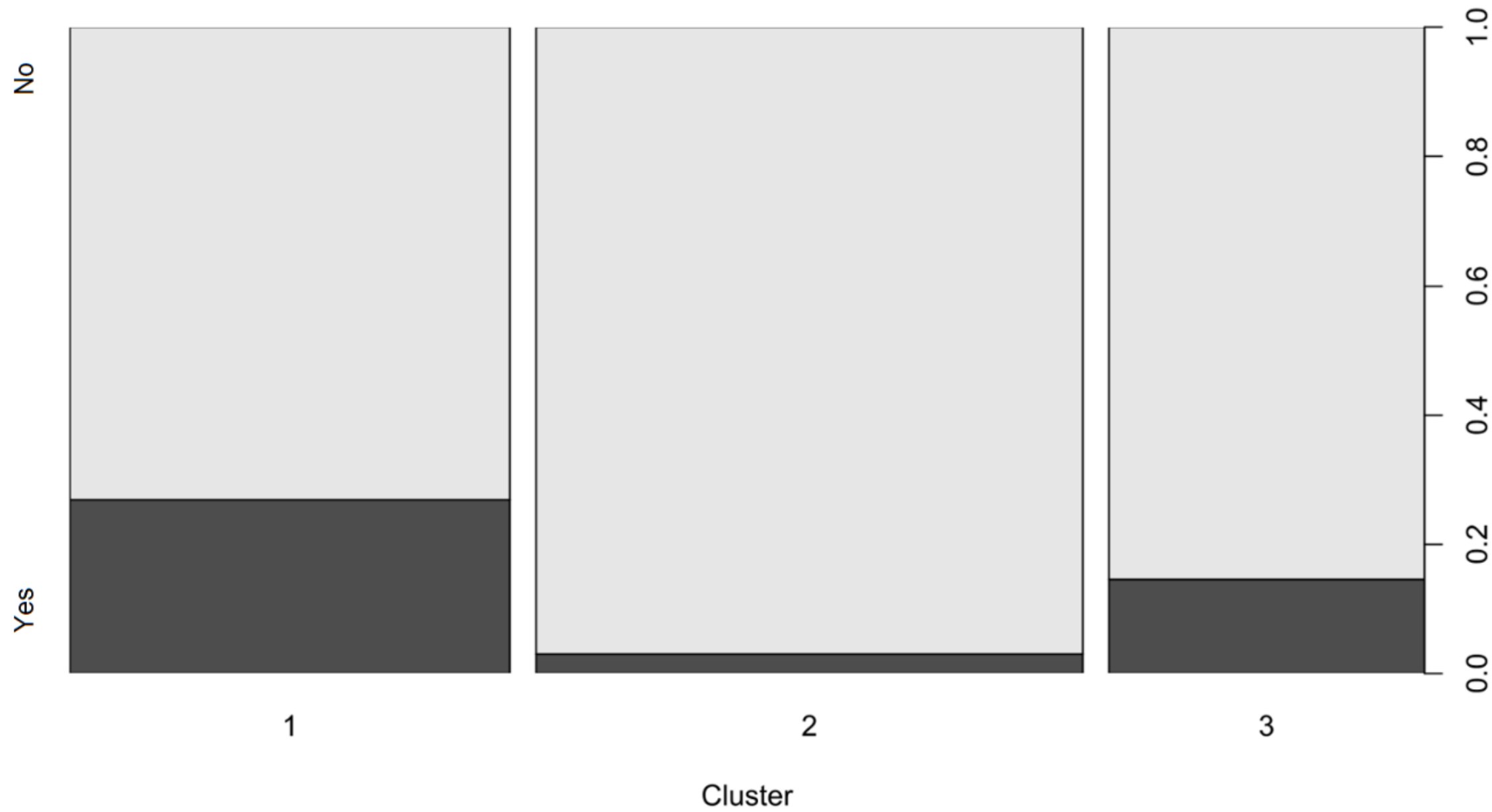

**Figure 7.** Mosaic plot between cluster and exports

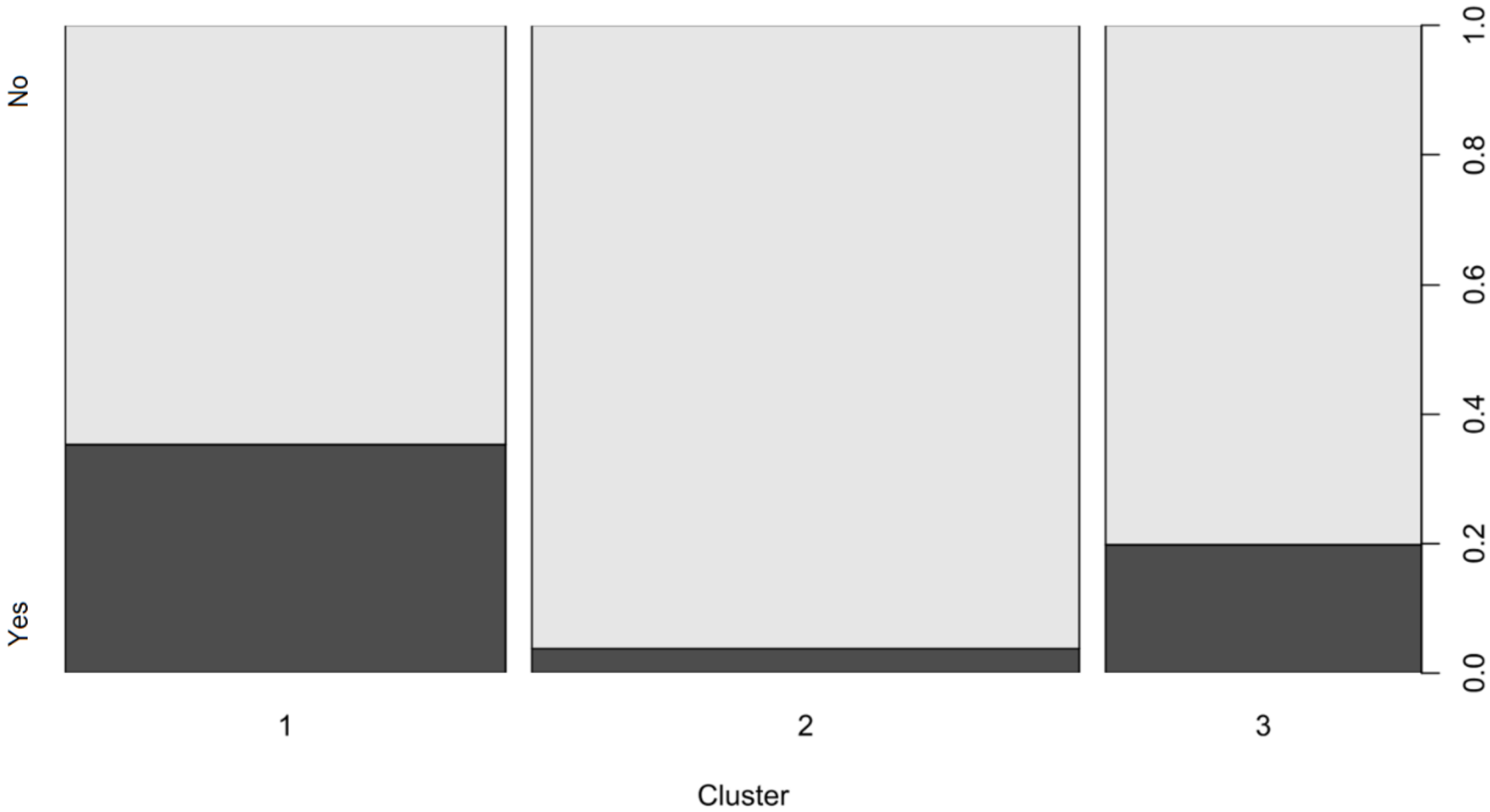

**Figure 8.** Mosaic plot between cluster and autonomous community (AN: Andalusia, AR: Aragon, CL: Castille-Leon, CM: Castille-la Mancha, CN: Cantabria, CT: Catalonia, GA: Galicia, MA: Madrid, MU: Murcia, OT: Other, PV: Basque country, VA: Valencian community)

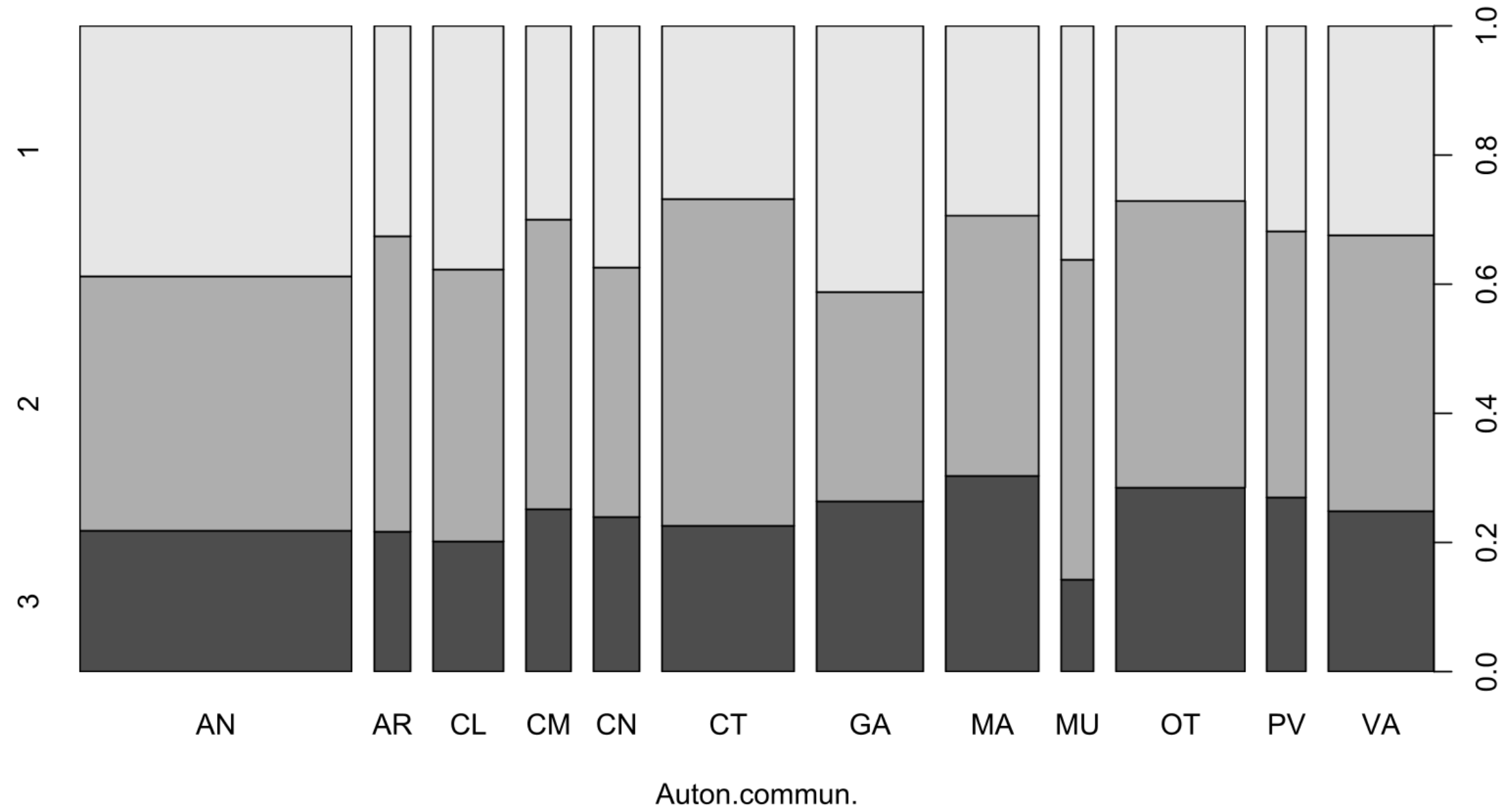

Figure 8 shows the importance of each cluster by regions (autonomous communities in Spain). The least profitable cluster 1 tends to concentrate in Andalusia, Castille-Leon, Cantabria and Galicia. Catalonia stands out for the most indebted cluster 2. The more resilient cluster 3 appears in Madrid and in the other communities (communities with fewer than 40 firms in the sample were grouped together).

## 5. Conclusion and discussion

The analysis of financial ratios reveals that, although firms in some of the sectors experienced financial strain in 2022 with a drop in profitability measures, they demonstrated remarkable resilience, leading to recovery and even surpassing pre-war performance levels by 2023. This recovery indicates the sector's ability to adapt and overcome the economic disruptions caused by the conflict.

The study also highlights the heterogeneity within the sector. While some sectors, such as the milling products and starches sector (106) and the cereal cultivation sector (0111) maintained or slightly improved their profitability, others, such as the bakery and farinaceous products sector (107), faced a pronounced deterioration in profitability in 2022. For instance, sector 106's rapid debt restructuring—facilitated by its smaller size—demonstrates the benefits of agile financial management.

A k-means cluster analysis of the standard financial ratios in Equations 2 to 13 resulted in a cluster containing 99.64 % of observations and two clusters containing only outliers (0.33 % and 0.03 % of observations respectively). Conversely, as in Arimany-Serrat and Coenders (2025a), Arimany-Serrat and Sgorla (2024), Dao et al. (2024), and Saus-Sala et al. (2024), compositional cluster analysis successfully reveals clusters of well-balanced sizes differing in resilience towards a crisis. The cluster analysis identified three distinct financial profiles: conservative (Cluster 1), high-risk leveraged (Cluster 2), and balanced (Cluster 3). Cluster 3 is the most resilient. Cluster 1 has less than ideal returns. Cluster 2’s overreliance on debt illustrates how aggressive leverage can magnify returns while also increasing the risk of insolvency. Unlike our expectations supported by similar research, Cluster 3 did not shrink during 2022. Its growth observed during 2023 does reinforce the idea that prolonged crises encourage firms to adopt more resilient strategies. This finding underscores the importance of diversified financing and adaptive governance to mitigate the impact of geopolitical disruptions.

Furthermore, the analysis revealed that cluster 1 is linked to the bakery sector (107), private firms, and high import/export activity, while Cluster 2 aligns with vegetable/animal oils (104), public limited companies, and low trade engagement. Cluster 3 lacks sectoral dominance, features smaller firms, and grew in the third year supporting a better adaptation of some firms to the conflict. This trajectory could signal that the sector may be better prepared for future crises, provided incentives for balanced strategies—such as those observed in Cluster 3—are sustained or strengthened. Some regional differences also emerged.

While this study focuses on financial resilience, broader systemic policies could amplify these adaptive capacities. For instance, Ben Hassen and El Bilali (2022) emphasize that building resilient food systems is critical to weathering disruptions, a principle echoed in the growth of Cluster 3’s diversified financial approaches. Similarly, Abay et al. (2023) propose policy measures such as maintaining open trade flows for agricultural inputs, avoiding export restrictions, and targeting subsidies to vulnerable stakeholders. While the focus of this study is not on these policies, understanding their context can provide valuable insights for strategic planning in the agricultural sector. These systemic considerations must be paired with critical reforms in corporate governance. Notably, the prevalence of public limited companies in cluster 2 raises critical questions about corporate governance priorities. The dominance of high-risk, debt-driven strategies among publicly traded firms may reflect a systemic short-termism driven by shareholder pressure to maximize returns, even at the expense of long-term sustainability. This aligns with the ethical dilemma inherent in shareholder primacy models: balancing profitability demands with the need for operational resilience, particularly in essential sectors such as food production.

The main limitations of this research include firstly the restricted sample derived from the SABI database, which only encompasses companies with a corporate structure, thereby excluding micro-enterprises and individual businesses common in Spain’s agricultural and food production sectors, as their financial data are not publicly disclosed. While the sample size of the four analysed subsectors is admittedly very different, it reflects the population structure for the aggregated results.

Secondly, the three-year timeframe (2021–2023) limits the assessment of long-term resilience trajectories. 2023 was the last available year of data at the time the article was submitted. Future research should remedy this limitation.

Thirdly, the study is descriptive. No causal inference can be made. Other relevant shocks during the study period could have affected the results at the same time as the Ukraine-Russia war, chief among them being the severe drought wave. The average rainfall in Spain for the 1991-2020 period was 649 mm., which dropped to a mere 514 in 2021-2022 and 592 in 2022-2023 (source: Spanish meteorological agency). The royal decree 4/2023 issued in May 2023 severely limited watering the fields and water supply to the farms, but for the sheer survival of trees and animals. Causal inference requires using statistical models in conjunction with the CoDa approach (Farreras-Noguer et al., 2025). For modelling panel data such as we have in this article, see for instance Carreras-Simó and Coenders (2021) and Escaramís and Arbussà (2025).

Fourthly, the study is limited to k-means clustering. Other clustering methods have successfully been adapted to CoDa and could be explored, including hierarchical clustering (Linares-Mustarós et al., 2018), weighted clustering (Jofre-Campuzano and Coenders, 2022), fuzzy clustering (Molas-Colomer et al., 2024) and latent-class analysis (Ferrer-Rosell et al., 2016).

Fifthly, the findings are context-specific to the analysed subsectors (cereal cultivation, oils/fats manufacturing, milling products, and bakery/farinaceous production) and should not be generalized to other sectors without further validation.

Finally, the exclusion of other accounting figures and financial statements such as the cash-flow statement narrows the financial analysis, omitting insights into liquidity dynamics critical for crisis adaptation (Arimany-Serrat et al., 2022).

## ACKNOWLEDGMENTS

This work was supported by the Spanish Ministry of Science, Innovation and Universities MCIN/AEI/10.13039/501100011033 and ERDF—a way of making Europe (grant number PID2021-123833OB-I00); the Spanish Ministry of Health (grant number CIBERCB06/02/1002); the Department of Research and Universities of Generalitat de Catalunya (grant number 2021SGR01197); and AGAUR and the Department of Climate Action, Food and Rural Agenda of Generalitat de Catalunya (grant number 2023-CLIMA-00037).

## AUTHOR CONTRIBUTIONS

Conceptualization: MHR and GC; methodology: GC; formal analysis: MHR and GC; data curation: MHR; writing—original draft: MHR; writing—review & editing: GC; Supervision: GC; funding acquisition: GC.

## CONFLICTS OF INTEREST

The authors declare no conflict of interest.

## DATA AVAILABILITY

The curated data are available from the corresponding author upon reasonable request.

## USE OF ARTIFICIAL INTELLIGENCE (AI)

AI tools were used for language editing only.

## REFERENCES

Abay, K. A., Breisinger, C., Glauber, J., Kurdi, S., Laborde, D., & Siddig, K. (2023). The Russia-Ukraine war: Implications for global and regional food security and potential policy responses. *Global Food Security*, 36, 100675. https://doi.org/10.1016/j.gfs.2023.100675

Aitchison, J. (1982). The statistical analysis of compositional data (with discussion). *Journal of the Royal Statistical Society Series B (Statistical Methodology)*, 44(2), 139-177. https://doi.org/10.1111/j.2517-6161.1982.tb01195.x

Aitchison, J. (1983). Principal component analysis of compositional data. *Biometrika*, 70(1), 57-65. https://doi.org/10.1093/biomet/70.1.57

Aitchison, J. (1986). *The statistical analysis of compositional data. Monographs on statistics and applied probability*. London, UK: Chapman and Hall.

Aitchison, J. (1997). The one-hour course in compositional data analysis or compositional data analysis is simple. In V. Pawlowsky-Glahn (Ed.), *Proceedings of the IAMG'97-The 3rd annual conference of the international association for mathematical geology*, (pp. 3-35). Barcelona, E: International Center for Numerical Methods in Engineering (CIMNE).

Aitchison, J., Barceló-Vidal, C., Martín-Fernández, J. A., & Pawlowsky-Glahn V. (2000). Logratio analysis and compositional distances. *Mathematical Geology*, 32(3), 271-275. https://doi.org/10.1023/A:1007529726302

Amat, O., & Manini, R. (2017). Credit scoring for the supermarket and retailing industry: analysis and application proposal. *European Accounting and Management Review*, 4(1), 75-88. https://doi.org/10.26595/eamr.2014.4.1.4

Arimany-Serrat, N., Carbonés-Trulls, L., Sabata-Aliberch, A., & Viladecans-Riera, C. (2017). Financial analysis of the feed sector in Catalonia. *European Accounting and Management Review,* 3(2), 89-107. https://doi.org/10.26595/eamr.2014.4.1.5

Arimany-Serrat, N., & Coenders, G. (2025). Biodiversity and accounting: Compositional methodology in the accounting statement analysis of the beekeeping industry. *Economía Agraria y Recursos Naturales*, 25(1), 5-34. https://doi.org/10.7201/earn.2025.01.01

Arimany-Serrat, N., Farreras Noguer, M. À., & Coenders, G. (2022). New developments in financial statement analysis: liquidity in the winery sector. *Accounting*, 8(3), 355-366. https://doi.org/10.5267/j.ac.2021.10.002

Arimany-Serrat, N., Farreras-Noguer, M. À., & Coenders, G. (2023). Financial resilience of Spanish wineries during the COVID-19 lockdown. *International Journal of Wine Business Research*, 35(2), 346-364. https://doi.org/10.1108/IJWBR-03-2022-0012

Arimany-Serrat, N., & Sgorla, A.F. (2024). Financial and ESG analysis of the beer sector pre and post COVID-19 in Italy and Spain. *Sustainability*, 16(17), 7412. https://doi.org/10.3390/su16177412

Baixauli-Soler, J. S., Lozano-Reina, G., Álvarez-Díez, S., & Rodríguez-Linares Rey, D. (2024). Impact of public guarantees on optimal debt levels following the COVID-19 pandemic: efficiency in their allocation. *Spanish Journal of Finance and Accounting/Revista Española de Financiación y Contabilidad*, 53(2), 176-202. https://doi.org/10.1080/02102412.2023.2177460

Ben Hassen, T., & El Bilali, H. (2022). Impacts of the Russia-Ukraine war on global food security: towards more sustainable and resilient food systems? *Foods*, 11(15), 2301. https://doi.org/10.3390/foods11152301

Bielieskov, M., & Szeligowski, D. (2024). Two years of Russian all-out aggression against Ukraine: State of play and future scenarios. OSCE Annual Security Review Conference. Vienna, AT.

Boungou, W., & Yatié, A. (2022). The impact of the Ukraine–Russia war on world stock market returns. *Economics Letters*, 215, 110516. https://doi.org/10.1016/j.econlet.2022.110516

Caliński, T., & Harabasz, J. (1974). A dendrite method for cluster analysis. *Communications in Statistics-Theory and Methods*, 3(1), 1-27. https://doi.org/10.1080/03610927408827101

Capece, G., Cricelli, L., Di Pillo, F., & Levialdi, N. (2010). A cluster analysis study based on profitability and financial indicators in the Italian gas retail market. *Energy Policy*, 38(7), 3394-3402.https://doi.org/10.1016/j.enpol.2010.02.013

Carreras-Simó, M., & Coenders, G. (2020). Principal component analysis of financial statements. A compositional approach. *Revista de Métodos Cuantitativos para la Economía y la Empresa*, 29, 18-37. https://doi.org/10.46661/revmetodoscuanteconempresa.3580

Carreras-Simó, M., & Coenders, G. (2021). The relationship between asset and capital structure: A compositional approach with panel vector autoregressive models. *Quantitative Finance and Economics*, 5(4), 571-590. https://doi.org/10.3934/QFE.2021025

Caruso, G., Gattone, S. A., Fortuna, F., & Di Battista, T. (2018). Cluster analysis as a decision-making tool: A methodological review. In E. Bucciarelli, S.-H. Chen, & J. M. Corchado (Eds.), *Decision economics: In the tradition of Herbert A. Simon's heritage*, (pp. 48-55). Cham, CH: Springer. https://doi.org/10.1007/978-3-319-60882-2_6

Cavero Rubio, J. A., Amoros Martinez, A., & Collazo Mazon, A. (2021). Economic effects of goodwill accounting practices: systematic amortisation versus impairment test. *Spanish Journal of Finance and Accounting/Revista Española de Financiación y Contabilidad*, 50(2), 224-245. https://doi.org/10.1080/02102412.2020.1778376

Chen, K. H., & Shimerda, T. A. (1981). An empirical analysis of useful financial ratios. *Financial Management*, 10(1), 51-60. https://doi.org/10.2307/3665113

Chiang, T. C. (2022). The effects of economic uncertainty, geopolitical risk and pandemic upheaval on gold prices. *Resources Policy*, 76, 102546. https://doi.org/10.1016/j.resourpol.2021.102546

Cirera, X., Cruz, M., Grover, A., Iacovone, L., Medvedev, D., Pereira-Lopez, M., & Reyes, S. (2021). Firm recovery during COVID-19: Six stylized facts. The World Bank. https://doi.org/10.1596/1813-9450-9810

Coenders, G. (2025). Application aux ratios financiers. In F. Bertrand, A. Gégout-Petit, & C. Thomas-Agnan (Eds.), *Données de composition,* (pp. 227-244). Paris, FR: Éditions TECHNIP.

Coenders, G., & Arimany-Serrat, N. (2023). Accounting statement analysis at industry level. A gentle introduction to the compositional approach. *arXiv*, 2305.16842. https://arxiv.org/abs/2305.16842

Coenders, G., Arimany-Serrat, N. (2025a). *Análisis estadístico de la información financiera. Metodología composicional*. Girona, ES: Documenta Universitaria. https://doi.org/10.33115/b/9788499847016

Coenders, G., & Arimany-Serrat, N. (2025b). Ten years of compositional analysis of financial ratios: Review, guidelines and future directions. *European Accounting and Management Review*, 11(1), 1–32. https://doi.org/10.36592/eamr.v11i1.01

Coenders, G., Egozcue, J. J., Fačevicová, K., Navarro-López, C., Palarea-Albaladejo, J., Pawlowsky-Glahn, V., & Tolosana-Delgado, R. (2023). 40 years after Aitchison's article "The statistical analysis of compositional data". Where we are and where we are heading. *SORT. Statistics and Operations Research Transactions*, 47(2), 207-228. https://doi.org/10.57645/20.8080.02.6

Comas-Cufí, M., & Thió-Henestrosa, S. (2011). CoDaPack 2.0: a stand-alone, multi-platform compositional software. In J. J. Egozcue, R. Tolosana-Delgado, & M. I. Ortego (Eds.), *CoDaWork'11: 4th international workshop on compositional data analysis. Sant Feliu de Guíxols*, (pp. 1-10). Girona, E: Universitat de Girona.

Cowen, S. S., & Hoffer, J. A. (1982). Usefulness of financial ratios in a single industry. *Journal of Business Research*, 10(1), 103-118. https://doi.org/10.1016/0148-2963(82)90020-0

Creixans-Tenas, J., Coenders, G., & Arimany-Serrat, N. (2019). Corporate social responsibility and financial profile of Spanish private hospitals. *Heliyon*, 5(10), e02623. https://doi.org/10.1016/j.heliyon.2019.e02623

Dao, B. T. T., Coenders, G., Lai, P. H., Dam, T. T. T., & Trinh, H. T. (2024). An empirical examination of financial performance and distress profiles during COVID-19: The case of fishery and food production firms in Vietnam. *Journal of Financial Reporting and Accounting*. https://doi.org/10.1108/JFRA-09-2023-0509

Deshpande, A. (2023). The effect of financial leverage on firm profitability and working capital management in the Asia-Pacific Region. *The Central European Review of Economics and Management*, 7(4), 43-71. https://doi.org/10.29015/cerem.977

Escaramís, G., & Arbussà, A. (2025). Considerations on the use of financial ratios in the study of family businesses. *arXiv*, 2501.16793. https://arxiv.org/abs/2501.16793

Escudero, A., Mancheño Losa, S., & López Pérez, J. J. (2022). Anuario de estadística (No. 2022, p. 159). Madrid, ES: Ministerio de Agricultura, Pesca y Alimentación, Gobierno de España.

European Commission, DG Agriculture and Rural Development. (2024). Short-term outlook for EU agricultural markets, Autumn 2024, 39, 34.

FAO. (2023). Crop prospects and food situation – Quarterly Global Report No. 2. FAO. https://doi.org/10.4060/cc6806en

Farreras-Noguer, M. À., Carreras-Pijuan, M., Linares-Mustarós, S., & Arimany-Serrat, N. (2025). Subsidies and financial performance in the rural tourism sector. *European Accounting and Management Review*, 11(1), 33–54. https://doi.org/10.36592/eamr.v11i1.02

Feranecová, A., & Krigovská, A. (2016). Measuring the performance of universities through cluster analysis and the use of financial ratio indexes. *Economics & Sociology*, 9(4), 259-271. https://doi.org/10.14254/2071-789X.2016/9-4/16

Ferrer-Rosell, B., Coenders, G., & Martínez-Garcia, E. (2016). Segmentation by tourist expenditure composition. An approach with compositional data analysis and latent classes. *Tourism Analysis*, 21(6), 589-602. https://doi.org/10.3727/108354216X14713487283075

Filzmoser, P., Hron, K., & Templ, M. (2018). *Applied compositional data analysis with worked examples in R*. New York, NY: Springer

Frecka, T. J., & Hopwood, W. S. (1983). The effects of outliers on the cross-sectional distributional properties of financial ratios. *The Accounting Review*, 58(1), 115-128.

Glauber, J. W., Laborde Debucquet, D., & Mamun, A. (2023). The impact of the Ukraine crisis on the global vegetable oil market. In J. W. Glauber, & D. Laborde Debucquet (Eds.), *The Russia-Ukraine conflict and global food security*, (pp. 33-37). Washington, DC: International Food Policy Research Institute (IFPRI).

Greenacre, M. (2018). *Compositional data analysis in practice*. New York, NY: Chapman and Hall/CRC press.

Iotti, M., Ferri, G., Manghi, E., Calugi, A., & Bonazzi, G. (2024a). Sustainability assessment of the performance of parmigiano reggiano PDO firms: A comparative analysis of firms' legal form and altitude range. *Sustainability*,16(20), 9093. https://doi.org/10.3390/su16209093

Iotti, M., Manghi, E., & Bonazzi, G. (2024b). Debt sustainability assessment in the biogas sector: application of interest coverage ratios in a sample of agricultural firms in Italy. *Energies*, 17(6), 1404. https://doi.org/10.3390/en17061404

Jagtap, S., Trollman, H., Trollman, F., Garcia-Garcia, G., Parra-López, C., Duong, L., Martindale, W., Munekata, P. E. S., Lorenzo, J. M., Hdaifeh, A., Hassoun, A., Salonitis, K., & Afy-Shararah, M. (2022). The Russia-Ukraine conflict: its implications for the global food supply chains. *Foods*, 11(14), 2098. https://doi.org/10.3390/foods11142098

Jofre-Campuzano, P., Coenders, G. (2022). Compositional classification of financial statement profiles. The weighted case. *Journal of Risk and Financial Management*, 15(12), 546. https://doi.org/10.3390/jrfm15120546

Kaufman, L., & Rousseeuw, P. J. (1990). *Finding groups in data: An introduction to cluster analysis*. New York, NY: Wiley.

Krylov, S. (2018). Target financial forecasting as an instrument to improve company financial health. *Cogent Business & Management*, 5(1), 1540074. https://doi.org/10.1080/23311975.2018.1540074

Kong, T., Yang, X., Wang, R., Cheng, Z., Ren, C., Liu, S., Li, Z., Wang, F., Ma, X., & Zhang, X. (2022). One year after COVID: the challenges and outlook of Chinese micro-and-small enterprises. *China Economic Journal*, 15(1), 1-28. https://doi.org/10.1080/17538963.2021.1995246

Lev, B., & Sunder, S. (1979). Methodological issues in the use of financial ratios. *Journal of Accounting and Economics*, 1(3), 187-210. https://doi.org/10.1016/0165-4101(79)90007-7

Linares-Mustarós, S., Coenders, G., & Vives-Mestres, M. (2018). Financial performance and distress profiles. From classification according to financial ratios to compositional classification. *Advances in Accounting*, 40, 1-10. https://doi.org/10.1016/j.adiac.2017.10.003

MacQueen, J. (1967). Some methods for classification and analysis of multivariate observations. In L. Lecam, & J. Neyman (Eds), *Proceedings of the fifth Berkeley symposium on mathematical statistics and probability Vol.1*, (pp. 281-297). Berkeley, CA: University of California Press.

Magrini, A. (2025). Bankruptcy risk prediction: a new approach based on compositional analysis of financial statements. *Big Data Research*, 41, 1000537. https://doi.org/10.1016/j.bdr.2025.100537

Malik, M., Kravchenko, S., Shpykuliak, O., & Hudz, H. (2024). Development of small businesses producing cereals, legumes, and sunflower seeds in wartime conditions. *Ekonomika APK*, 31(1), 41-53. https://doi.org/10.32317/2221-1055.202401041

Martín-Fernández, J. A., Palarea-Albaladejo, J., & Olea, R. A. (2011). Dealing with zeros. In V. Pawlowsky-Glahn, & A. Buccianti (Eds.), *Compositional data analysis. Theory and applications*, (pp. 47-62). New York, NY: Wiley.

Molas-Colomer, X., Linares-Mustarós, S., Farreras-Noguer, M. À., & Ferrer-Comalat, J. C. (2024). A new methodological proposal for classifying firms according to the similarity of their financial structures based on combining compositionaldata with fuzzy clustering. *Journal of Multiple-Valued Logic and Soft Computing*, 43(1-2), 73-100.

Palarea-Albaladejo, J., & Martín-Fernández, J. A. (2008). A modified EM alr-algorithm for replacing rounded zeros in compositional data sets. *Computers & Geosciences*, 34(8), 902-917.https://doi.org/10.1016/j.cageo.2007.09.015

Pawlowsky-Glahn, V., Egozcue, J. J., & Tolosana-Delgado, R. (2015). *Modeling and analysis of compositional data*. Chichester, UK: Wiley.

Saleh, I., Marei, Y., Ayoush, M., & Abu Afifa, M. M. (2023). Big data analytics and financial reporting quality: qualitative evidence from Canada. *Journal of Financial Reporting and Accounting*, 21(1), 83-104. https://doi.org/10.1108/JFRA-12-2021-0489

Saus-Sala, E., Farreras-Noguer, M. À., Arimany-Serrat N., Coenders, G. (2021). Compositional DuPont analysis. A visual tool for strategic financial performance assessment. In P. Filzmoser, K. Hron, J. A. Martín-Fernández, & J. Palarea-Albaladejo (Eds.), *Advances in compositional data analysis. Festschrift in honour of Vera Pawlowsky-Glahn*, (pp. 189-206). Cham, CH: Springer.

Saus-Sala, E., Farreras Noguer, M. À., Arimany Serrat, N., & Coenders, G. (2023). Análisis de las empresas de turismo rural en Cataluña y Galicia: Rentabilidad económica y solvencia 2014 – 2018. *Cuadernos del CIMBAGE*, 1(25), 33-54. https://doi.org/10.56503/CIMBAGE/Vol.1/Nro.25(2023)p.33-54

Saus-Sala, E., Farreras-Noguer, M. À., Arimany-Serrat, N., & Coenders, G. (2024). Financial analysis of rural tourism in Catalonia and Galicia pre- and post COVID-19. *International Journal of Tourism Research*, 26(4), e2698. https://doi.org/10.1002/jtr.2698

Sharif, A., Aloui, C., & Yarovaya, L. (2020). COVID-19 pandemic, oil prices, stock market, geopolitical risk and policy uncertainty nexus in the US economy: Fresh evidence from the wavelet-based approach. *International Review of Financial Analysis*, 70, 101496. https://doi.org/10.1016/j.irfa.2020.101496

Sharma, S., Shebalkov, M., & Yukhanaev, A. (2016). Evaluating banks performance using key financial indicators - a quantitative modeling of Russian banks. *The Journal of Developing Areas*, 50(1), 425-453. https://doi.org/10.1353/jda.2016.0015

Tascón, M. T., Castaño, F. J., & Castro, P. (2018). A new tool for failure analysis in small firms: frontiers of financial ratios based on percentile differences (PDFR). *Spanish Journal of Finance and Accounting/Revista Española de Financiación y Contabilidad*, 47(4), 433-463. https://doi.org/10.1080/02102412.2018.1468058

Teixeira da Silva, J., Koblianska, I., & Kucher, A. (2023). Agricultural production in Ukraine: An insight into the impact of the Russo-Ukrainian war on local, regional and global food security. *Journal of Agricultural Sciences*, 68(2), 121-140. https://doi.org/10.2298/JAS2302121T

Thió-Henestrosa, S., & Martín-Fernández, J. A. (2005). Dealing with compositional data: The freeware CoDaPack. *Mathematical Geology*, 37(7), 773-793. https://doi.org/10.1007/s11004-005-7379-3

Van den Boogaart, K. G., & Tolosana-Delgado, R. (2013). *Analyzing compositional data with R*. Berlin, DE: Springer. https://doi.org/10.1007/978-3-642-36809-7

Ward jr, J. H. (1963). Hierarchical grouping to optimize an objective function. *Journal of the American Statistical Association*, 58(301), 236-244. https://doi.org/10.1080/01621459.1963.10500845